\documentclass[conference]{IEEEtran}
\IEEEoverridecommandlockouts
\usepackage{amssymb}
\usepackage{amsfonts}

\usepackage{enumerate}

\usepackage{enumerate}
\usepackage{amsmath,amsthm}
\usepackage{mathtools}
\usepackage{algorithm,algorithmic}
\usepackage{float}
\usepackage{hyperref}
\usepackage{color}
\usepackage{makeidx}
\usepackage{bbm}
\usepackage{graphicx}
\usepackage{lipsum}
\usepackage{soul}
\usepackage{tabularx}
\usepackage{dsfont}
\usepackage[table,xcdraw]{xcolor}

\usepackage{amsfonts}
\usepackage{times}
\usepackage{graphicx}
\usepackage{latexsym}
\usepackage{dsfont}
\usepackage{amssymb}
\usepackage{amsmath}
\usepackage{cite}
\usepackage{verbatim}
\usepackage{subfigure}

\def\bb0{{\mathbb{0}}}

\def\bb{{\mathbf{b}}}

\def\bh{{\mathbf{h}}}

\def\bw{{\mathbf{w}}}

\def\b0{{\mathbf{0}}}

\def\sf0{{\mathsf{0}}}

\usepackage{epstopdf}

\newcommand{\sref}[1]{{Section}~\ref{#1}}
\newcommand{\fref}[1]{{Fig.}~\ref{#1}}

\newcommand{\tref}[1]{{Table}~\ref{#1}}

\DeclareMathOperator*{\argmax}{arg\,max}

\newcommand{\subto}{\operatorname{s.t.}}

\begin{document}
\title{A Digital Twin Assisted Framework for Interference Nulling in Millimeter Wave MIMO Systems}
\author{Yu Zhang, Tawfik Osman, and Ahmed Alkhateeb \thanks{Yu Zhang, Tawfik Osman, and Ahmed Alkhateeb are with Arizona State University (Email: y.zhang, tmosman, alkhateeb@asu.edu). This work is supported in part by the National Science Foundation under Grant No. 1923676 and by a U. S. Army research program under contract No. W911NF21C0015.}}
\maketitle

\begin{abstract}

Millimeter wave (mmWave) and terahertz MIMO systems rely on pre-defined beamforming codebooks for both initial access and data transmission. However, most of the existing codebooks adopt pre-defined beams that focus mainly on improving the gain of their target users, without taking interference into account, which could incur critical performance degradation in the dense networks. To address this problem, in this paper, we propose a sample-efficient digital twin-assisted beam pattern design framework that learns how to form the beam pattern to reject the signals from the interfering directions. The proposed approach does not require any explicit channel knowledge or any coordination with the interferers. The adoption of the digital twin improves the sample efficiency by better leveraging the underlying signal relationship and by incorporating a demand-based data acquisition strategy. Simulation results show that the developed signal model-based learning framework can significantly reduce the actual interaction with the radio environment (i.e., number of measurements) compared to the model-unaware design, leading to a more practical and efficient interference-aware beam design approach.

\end{abstract}

\section{Introduction} \label{intro}

Rejecting in-band interference in the RF domain of the hybrid or fully analog MIMO systems is possible owing to the extra degree of freedom brought by multiple antennas.
However, realizing its full potential is challenging in practical systems due to the lack of channel knowledge as well as the additional hardware constraints.
As a result, the existing approaches normally require large beam learning overhead in order to form well-shaped analog beams.
Moreover, when the codebook design problem is considered, such learning overhead gets magnified and increases linearly with respect to the codebook size, since the learning experience of one beam is not transferable to the others.
Such prohibitive overhead downplays the potential of the system, and it motivates the development of sample efficient interference-aware beam codebook design framework, which is the focus of this paper.

\textbf{Contributions:}
In this paper, we develop a model-based digital twin-assisted learning framework that achieves higher sample efficiency by better leveraging the underlying signal model. The improvement on the sample efficiency has the potential of reducing the beam learning overhead as well as shortening the convergence time of the proposed solution. The proposed digital twin-assisted learning framework also provides flexibility to the practical deployment in terms of enabling data sharing and cooperation (hence better learning) in complex scenarios.
We extensively evaluate the proposed interference-aware beam learning framework using numerical simulations. This provides a comprehensive assessment of the capability of the proposed approach in nulling interference without requiring any knowledge about the channel, array geometry, or user location.
The role of the digital twin model is also tested which empirically shows its efficacy in guiding the beam learning process and highlights its potential gain of reducing the overall learning overhead.

\textbf{Prior work:}
The prior work of interference nulling beamforming design either focuses on MIMO transceivers with fully-digital architectures \cite{Lorenz2005,Gesbert2010,Dahrouj2010},
or assumes some kind of channel information, such as covariance, of both the target and interferers \cite{Smith1999Optimum,Luyen2017}.
Some other approaches that do not rely on channel information are also proposed in the past \cite{Steyskal1983,Davis1998}. But they, in general, do not leverage underlying signal relationship which leads to large beam learning overhead and are not robust.

\section{System and Channel Models} \label{sec:System}

We consider the communication system where a mmWave MIMO base station (BS), equipped with $M$ antennas, is communicating with a single-antenna user equipment (UE) in an {uplink mode}.
Moreover, we assume that there exist $K$ ($\ge 1$) non-cooperative interference transmitters\footnote{For ease of exposition, each interference transmitter is also assumed to have a single antenna. This means that the interference signals are being transmitted omni-directionally.} in the vicinity of the BS, operating at the same frequency bands and hence causing interference to the BS receiver.
Therefore, if the UE transmits a symbol $x\in\mathbb{C}$ to the BS, and the other $K$ interference transmitters also transmit symbols $x_k\in\mathbb{C}, k=1,\dots,K$ at the same time and frequency slot, such that all the transmitted symbols satisfy the same average power constraint, i.e., $\mathbb{E}[|x|^2]=P_x$ and $\mathbb{E}[|x_k|^2]=P_x, \forall k$, the received signal at the BS after combining can then be expressed as
\begin{equation}\label{rec}
  y = {\bf w}^H{\bf h} x + \sum_{k=1}^{K}{\bf w}^H{\bf h}_k x_k + {\bf w}^H{\bf n},
\end{equation}
where ${\bf h}\in\mathbb{C}^{M\times 1}$ is the channel between the BS and the UE, ${\bf h}_k\in\mathbb{C}^{M\times 1}$ is the channel between the BS and the $k$-th interference transmitter.
It is worth pointing out here that for clarity, we subsume the factors such as path-loss and transmission power into the channels.
${\bf n} \sim \mathcal{CN}(0, \sigma^2{\bf I})$ is the receive noise vector at the BS with $\sigma^2$ being the noise power and ${\bf w}\in\mathbb{C}^{M\times 1}$ is the combining vector used by the BS.
Furthermore, given the high cost and power consumption of the mixed-signal components, we consider a practical system where the BS has only one radio frequency (RF) chain\footnote{It is very important to note that the RF precoder in a system with hybrid architecture is normally constructed using pre-defined codebooks that have pre-determined beams. Therefore, the learned beams in this paper can be included in such codebooks and be used in the hybrid analog/digital architectures as well.} and employs analog-only beamforming/combining using a network of $r$-bit quantized phase shifters.
Therefore, the combining vector at the BS can be written as
\begin{equation}\label{Analog}
  {\bf w} = \frac{1}{\sqrt{M}}\left[ e^{j\theta_1}, e^{j\theta_2}, \dots, e^{j\theta_M} \right]^T,
\end{equation}
where each phase shift $\theta_m, \forall m=1,\dots,M$ is selected from a finite set $\boldsymbol{\Psi}$ with $2^r$ possible discrete values drawn uniformly from $(-\pi, \pi]$. The normalization factor $M^{-1/2}$ is to make sure the combiner has unit power, i.e., $\|\mathbf{w}\|_2^2=1$.

We adopt a  geometric channel model for the channel between BS and UE, as well as the interference channels between BS and any interfering transmitters.
Hence, the channel between BS and its served UE takes the following form (the channel between BS and any interference transmitter takes similar form)
\begin{equation}\label{ch}
  {\bf h} = \sum_{\ell=1}^{L} \alpha_{\ell}{\bf a}(\phi_{\ell}, \vartheta_{\ell}),
\end{equation}
where $L$ is the number of multi-paths.
Each path $\ell$ has a complex gain $\alpha_{\ell}$, which includes the  path-loss.
The angles $\phi_{\ell}$ and $\vartheta_{\ell}$ represent the $\ell$-th path's azimuth and elevation angles of arrival respectively, and ${\bf a}(\phi_{\ell}, \vartheta_{\ell})$ is the BS array response vector.
The exact expression of ${\bf a}(\phi_{\ell}, \vartheta_{\ell})$ depends on the array geometry and possible hardware impairments.

\section{Problem Formulation} \label{sec:special}

In this paper, we investigate the design of the analog combining/precoding that achieves interference awareness (i.e., attempts to address the interference) without explicitly knowing any channel state information.
Given the receive signal \eqref{rec} at the BS, the achievable rate of its target user can be written as
\begin{equation}\label{ach}
  R = \log_2\left(1 + \frac{|{\bf w}^H\mathbf{h}|^2P_x}{\sum_{k=1}^K|{\bf w}^H\mathbf{h}_{k}|^2P_x + \sigma^2}\right).
\end{equation}
The objective is to design the combining vector ${\bf w}$ such that the achievable rate of the target user, i.e., \eqref{ach}, can be maximized.
Given the monotonicity of the logarithm function, this is equivalent to maximize the SINR term in \eqref{ach}.
Therefore, the problem can be cast as
\begin{align}\label{prob}
 {\bf w}^\star = \argmax\limits_{{\bf w}} & \hspace{2pt}  \frac{|{\bf w}^H\mathbf{h}|^2P_x}{\sum_{k=1}^K|{\bf w}^H\mathbf{h}_{k}|^2P_x + \sigma^2}, \\
 \subto  \hspace{2pt} &  w_m = \frac{1}{\sqrt{M}}e^{j\theta_{m}}, ~ \forall m=1, ..., M, \label{cons-1} \\
 & \theta_{m}\in\boldsymbol{\Psi}, ~ \forall m=1, ..., M, \label{cons-2}
\end{align}
where $w_m$ is the $m$-th element of the combining vector ${\bf w}$ \footnote{It is important to note that the proposed interference-aware beam  learning approach can be straightforwardly extended to learning a codebook with multiple beams by, for example, using the user clustering and assignment algorithm proposed in \cite{Zhang2022Reinforcement}.}.
Solving the interference-aware beam pattern design problem formulated in \eqref{prob} is challenging due to the non-convex and discrete hardware constraints \eqref{cons-1} and \eqref{cons-2}, as well as the unknown channels in the objective function \eqref{prob}.
Therefore, it is hard to solve \eqref{prob} using the conventional optimization methods \cite{Lorenz2005,Gesbert2010,Dahrouj2010}. An important observation, however, is that for a given combining beam $\bw$, evaluating the SINR requires only the power values (after combining) of the desired and interference signals, and does not require explicit knowledge about the channel vectors. 
With this observation, we cast our problem as developing an online machine learning approach that learns how to design an interference-aware beam pattern $\bw$ that optimizes \eqref{prob}, \textbf{given only the receive power measurements} for the signal plus interference and noise, $\left|\bw^H \bh\right|^2P_x + \sum_{k=1}^K \left|\bw^H \bh_{k}\right|^2P_x + \sigma^2$, and the interference plus noise, $\sum_{k=1}^K \left|\bw^H \bh_{k}\right|^2P_x + \sigma^2$.

\section{Online Learning of Interference Aware Beam Pattern Design} \label{sec:BPL}

In this section, we present the proposed online reinforcement learning based interference-aware beam pattern learning approach. To solve the problem with reinforcement learning, we first fit all the ingredients of problem \eqref{prob} into a general reinforcement learning framework as follows:

\begin{itemize}
  \item \textbf{State:} We define the state ${\bf s}_t$ as a vector that consists of the phases of all the phase shifters at the $t$-th iteration, that is, ${\bf s}_t=\left[\theta_1, \theta_2, \dots, \theta_M\right]^T$.
  \item \textbf{Action:} We define the action ${\bf a}_t$ as the element-wise changes to all the phases in ${\bf s}_t$. Since the phases can only take values in $\boldsymbol\Psi$, a change of a phase represents the action that a phase shifter selects a value from $\boldsymbol\Psi$. Therefore, the action is directly specified as the next state, i.e., ${\bf a}_t = {\bf s}_{t+1}$, which can be viewed as a deterministic transition in the Markov Decision Process (MDP).
  \item \textbf{Reward:} We define a binary reward mechanism, i.e., the reward $r_t$ takes values from $\{+1, -1\}$. Since the objective of \eqref{prob} is to maximize the SINR performance, we compare the SINR achieved by the current combining vector, denoted as $\mathrm{SINR}_t$, with the previous one, i.e., $\mathrm{SINR}_{t-1}$. The reward is determined according to the following rule: $r_t=+1$, if $\mathrm{SINR}_t > \mathrm{SINR}_{t-1}$; $r_t=-1$, otherwise.
\end{itemize}

\noindent\textbf{Deep Reinforcement Learning Architecture}:
Given the reinforcement learning formulation above for the interference-aware beam learning problem, we adopt an actor-critic based deep reinforcement learning architecture.
This follows the learning framework that we proposed earlier in \cite{Zhang2022Reinforcement}. In summary, both the actor and critic networks are implemented using elegant fully-connected (FC) feed-forward neural networks.
The input of the actor network is the state and the output is the action, while the critic network takes in the state-action pair and outputs the predicted Q value.\footnote{The detailed architectures and the parameters of the adopted neural networks are provided in \sref{sub:DLA}.}
Moreover, to respect the discrete phase shifter hardware constraint \eqref{cons-2}, we perform an element-wise quantization to make the predicted action a valid one. To be more specific, assume that $\widehat{{\bf a}}_t$ is the predicted action from the actor network at time $t$. Then, the action that finally gets implemented to the system is given by
\begin{equation}\label{Quant}
  [{\bf a}_t]_m = \mathop{\arg\min}_{\theta\in\boldsymbol{\Psi}}\left|[\widehat{{\bf a}}_t]_m - \theta\right|, ~ \forall m=1, \dots, M.
\end{equation}
%
It is worth emphasizing that such quantization operation is only activated when the system is actually implementing the predicted action by the actor network to obtain reward.
It is not involved in the training process of the actor network due to its non-differentiability.

Despite its full compatibility with the considered system, the proposed interference-aware beam learning solution still has two  drawbacks.
First, it  requires a relatively large number of iterations to find a qualified beam pattern, especially when the number of antennas is large. As a result, this incurs a large beam learning overhead, since these iterations are done over the air.
Second, as indicated by the objective function of \eqref{prob}, the SINR performance of a given beam is determined by two factors: (i) The desired beamforming gain and (ii) The effectiveness of suppressing the undesired interference.
However, the proposed solution does not fully leverage this information as it only focuses on the overall SINR performance.
It turns out that the decomposition of these two factors, as will be further discussed in the next section, makes the data sharing among the learning processes of different beams possible, which has the potential of improving the convergence behavior of the beam/codebook learning algorithm.

\section{Digital Twin Assisted Beam Learning Framework} \label{sec:surrogate}

In this section, we describe in detail the proposed digital twin assisted interference-aware beam pattern learning framework.
The motivations of introducing the digital twin are mainly two-folds.
First, it has the potential of improving the sample efficiency (i.e., reducing the number of interactions with the actual environment) of the learning process \cite{DT_magazine}.
Second, it facilitates other more complex tasks (than learning a single beamforming vector), such as data sharing (which can be very useful in learning \emph{interference-aware beam codebooks}) and cooperative learning\footnote{For instance, as the system has full knowledge of its simulated environment, it can assign accurate reward to each agent. This has the potential of mitigating the non-stationary environment problem that exists in most of the multi-agent learning tasks.} (among multiple BSs to avoid interfering each other).

\subsection{Digital Twin for Beam Pattern Learning} \label{subsec:sur}

In this subsection, we introduce the proposed digital twin that assists the learning of interference-aware beams.
As mentioned before, in order to acquire the reward signal that is used for training the RL agent, the system needs to estimate two quantities, i.e., the signal power, $P_{\mathrm{S}} = \left|\bw^H \bh\right|^2P_x$, and the interference plus noise power, $P_{\mathrm{I+N}} = \sum_{k=1}^K \left|\bw^H \bh_{k}\right|^2P_x + \sigma^2$.
Therefore, correspondingly, there are two major components in the considered digital twin that provide the agent with such information, i.e., an interference predictor and a signal predictor, as will be discussed in this subsection.

\subsubsection{The key idea of digital twin}

The machine learning model that virtually interacts with the agent can be considered as a \textbf{digital twin}.
This model is used to \emph{imitate} the behavior of the actual environment, aiming to reduce the expensive (sometimes, even impossible) actual evaluations of the design.
In this paper, we design the digital twin with a particular emphasis on two important aspects.
\textbf{First,} a digital twin should be able to accurately model the behavior of the actual environment, i.e., having accurate predictions.
\textbf{Second,} training a digital twin should, in general, require less actual data samples than directly interacting with the actual environment, which yields a high sample efficiency.
With these important criterions in mind, we next describe the adopted digital twin.
As mentioned before, the considered digital twin consists of two major components, i.e., an interference prediction model and a signal prediction model.
Formally, the interference prediction model predicts the interference plus noise power based on the configuration of the receive combining vector, which can be expressed as
\begin{equation}\label{sur-int}
  \widehat{P}_{\mathrm{I+N}} = f_{\mathrm{in}}(\mathbf{w}; \boldsymbol{\Theta}_{\mathrm{in}}),
\end{equation}
where $\mathbf{w}\in\mathbb{C}^{M\times 1}$ is the input of the model, representing the designed receive combining vector, and the output is the predicted interference plus noise power, i.e., $\widehat{P}_{\mathrm{I+N}}\in\mathbb{R}$. The model is parameterized by $\boldsymbol{\Theta}_{\mathrm{in}}$.
Similarly, the signal prediction model predicts the signal power of a given receive combining vector, which can be written as
\begin{equation}\label{sur-sig}
  \widehat{P}_{\mathrm{S}} = f_{\mathrm{s}}(\mathbf{w}; \boldsymbol{\Theta}_{\mathrm{s}}),
\end{equation}
where $\widehat{P}_{\mathrm{S}}\in\mathbb{R}$ is the predicted signal power value and $\boldsymbol{\Theta}_{\mathrm{s}}$ denotes the model parameters.
It is worth mentioning that the architecture of $f_{\mathrm{in}}$ and $f_{\mathrm{s}}$ is not unique and is a design choice.
Next, we present two candidates that could be used in the considered beam learning task.

\subsubsection{Digital twin architecture}

In this paper, we study two specific designs: (i) A model-based prediction architecture, and (ii) a fully-connected neural network based prediction architecture.

\noindent\textbf{Model-based architecture:}
The model-based architecture, as its name suggests, is inspired by the underlying \emph{signal model}.
For instance, as can be seen from the expression of the interference plus noise power, i.e., $P_{\mathrm{I+N}} = \sum_{k=1}^K \left|\bw^H \bh_{k}\right|^2P_x + \sigma^2$, it takes a quadratic form of the receive combining vector $\bw$.
To see this, by defining $\mathbf{H}=[\mathbf{h}_1, \mathbf{h}_2, \dots, \mathbf{h}_K]$, $P_{\mathrm{I+N}}$ can be expressed as
\begin{align}\label{in-pwr}
  P_{\mathrm{I+N}} & = \left\|\mathbf{H}^H\mathbf{w}\right\|_2^2P_x + \sigma^2, \\
    & = \mathbf{w}^H\left(P_x\mathbf{H}\mathbf{H}^H + \sigma^2\mathbf{I}\right)\mathbf{w}, \\
    & = \mathbf{w}^H\mathbf{A}\mathbf{w}, \label{in-pwr-quadratic}
\end{align}
where $\mathbf{A} = P_x\mathbf{H}\mathbf{H}^H + \sigma^2\mathbf{I}$.
The signal power can be expressed in the similar form, i.e., $P_{\mathrm{S}} = \mathbf{w}^HP_x\mathbf{h}\mathbf{h}^H\mathbf{w}$.
Therefore, the interference prediction network is essentially leveraged to learn the relationship \eqref{in-pwr-quadratic}.
Inspired by this, we design the interference prediction network with a focus on imitating the ``behavior'' of $\mathbf{A}$.
Specifically, the interference prediction network is chosen to take the following form
\begin{equation}\label{in-net}
  f_{\mathrm{in}}(\mathbf{w}) = \mathbf{w}^H\mathbf{Q}_{\mathrm{in}}\mathbf{Q}_{\mathrm{in}}^H\mathbf{w},
\end{equation}
where $\mathbf{Q}_{\mathrm{in}}\in\mathbb{C}^{M\times r_{\mathrm{in}}}$ with $r_{\mathrm{in}}$ being a hyperparameter.
Therefore, the parameter of the interference prediction network is essentially $\mathbf{Q}_{\mathrm{in}}$, i.e., $\boldsymbol{\Theta}_{\mathrm{in}} = \{\mathbf{Q}_{\mathrm{in}}\}$.
The signal prediction network takes the similar form, i.e., $f_{\mathrm{s}}(\mathbf{w}) = \mathbf{w}^H\mathbf{Q}_{\mathrm{s}}\mathbf{Q}_{\mathrm{s}}^H\mathbf{w}$, where $\mathbf{Q}_{\mathrm{s}}\in\mathbb{C}^{M\times r_{\mathrm{s}}}$ with $r_{\mathrm{s}}$ being a hyperparameter as well, which makes $\boldsymbol{\Theta}_{\mathrm{s}} = \{\mathbf{Q}_{\mathrm{s}}\}$.

\noindent\textbf{Fully-connected neural network based architecture:}
Despite being lightweight and a better fit to the signal model, the model-based architecture, fundamentally, suffers from any mismatch between the assumed signal model and the actual signal relationship.
For instance, there are normally unknown non-linearities in the practical hardware that undermine the validity of the assumed relationship between the receive combining vector and the interference plus noise power (similarly for the signal power).
As a result, the signal model cannot always be met and the model-based architecture will show up certain level of residual prediction errors that are very hard to be eliminated given the less powerful expressive capability of its architecture.
Motivated by this, we also investigate a more general architecture, which is built upon fully-connected neural network, given its powerful universal approximation capability \cite{NNUnivApprox}.
Specifically, both $f_{\mathrm{in}}$ and $f_{\mathrm{s}}$ are modeled with feed-forward fully-connected neural networks.
The detailed network parameters will be provided in \sref{sub:DLA}.

\subsubsection{Training dataset and loss function}

We denote the training dataset of the interference prediction network as
\begin{equation}\label{in-data}
  \mathcal{D}_{\mathrm{in}} = \left\{\left(\mathbf{w}^{(n)}, P_{\mathrm{I+N}}^{(n)}\right)_{n=1}^{N_{\mathrm{in}}}\right\},
\end{equation}
where each data sample is comprised of a combining vector and its corresponding interference plus noise power value obtained from the actual environment, i.e., from the real measurement.
$N_{\mathrm{in}}$ is the total number of data samples in the dataset, i.e., $|\mathcal{D}_{\mathrm{in}}|=N_{\mathrm{in}}$.
And the dataset used for training the signal prediction network can be similarly denoted as
\begin{equation}\label{s-data}
  \mathcal{D}_{\mathrm{s}} = \left\{\left(\mathbf{w}^{(n)}, P_{\mathrm{S}}^{(n)}\right)_{n=1}^{N_{\mathrm{s}}}\right\},
\end{equation}
with $N_{\mathrm{s}}$ being its size.
%
Since the target of these two networks is to predict the power values, we pose the learning problem as a regression problem conducted in a supervised fashion.
Furthermore, we employ mean squared error (MSE) as the training loss function.
Using the interference prediction network as an example, for the $n$-th data sample in $\mathcal{D}_{\mathrm{in}}$, the loss function is defined as
\begin{align}\label{loss}
  \mathcal{L}\left(P_{\mathrm{I+N}}^{(n)}, \widehat{P}_{\mathrm{I+N}}^{(n)}\right) & = \left|P_{\mathrm{I+N}}^{(n)} - \widehat{P}_{\mathrm{I+N}}^{(n)}\right|^2, \\
    & = \left|P_{\mathrm{I+N}}^{(n)} - f_{\mathrm{in}}(\mathbf{w}^{(n)}; \boldsymbol{\Theta}_{\mathrm{in}})\right|^2.
\end{align}
The loss function used for the signal prediction network is identical.

\begin{figure}[t]
	\centering
	\includegraphics[width=1\columnwidth]{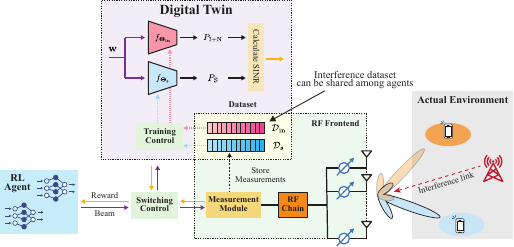}
    \caption{An illustration of the proposed digital twin-assisted interference-aware beam pattern learning framework.}
	\label{fig:surrogate-assisted}
\end{figure}

\subsection{Digital Twin Assisted Learning} \label{subsec:sur-ass}

In this subsection, we discuss how to integrate the digital twin with the proposed RL based beam learning framework.
Since the digital twin is essentially used to provide the RL agent with a simulated environment to interact with, it plays the same role as the actual environment.
However, in order to provide high quality \emph{synthetic} feedback, it requires training process that relies on the \emph{authentic} data collected from the actual environment.
Based on the trained digital twin, the system can virtually evaluate its designed beams without measuring the physical signals.
Moreover, the system might require constantly switching between the digital twin and the actual environment, triggered by the demand for the authentic data.
Next, we summarize the key components of the proposed digital twin assisted beam learning.

\noindent\textbf{Initial interaction and data acquisition:}
The system starts with the normal interaction between the RL agent and the actual environment.
To be more specific, upon forming a new beam $\tilde{{\bf w}}$, the BS estimates the interference plus noise power $P_{\mathrm{I+N}}$ and the signal power $P_{\mathrm{S}}$.
The reward signal used for RL agent learning will then be generated.
Moreover, these authentic power measurements together with the beam will be stored in the two datasets, i.e., $\mathcal{D}_{\mathrm{in}}$ and $\mathcal{D}_{\mathrm{s}}$, respectively.
During this interaction process, two initial datasets are established.

\noindent\textbf{Digital twin training:}
Based on the collected initial datasets $\mathcal{D}_{\mathrm{in}}$ and $\mathcal{D}_{\mathrm{s}}$, the two sub-networks of the digital twin, i.e., the interference prediction network $f_{\mathrm{in}}$ and the signal prediction network $f_{\mathrm{s}}$, are trained in a \emph{supervised} manner.
After the training process saturates, the digital twin is ready to interact with the RL agent.

\noindent\textbf{Environment switching and virtual interaction:}
The switching from the actual environment to the digital twin is triggered based on multiple factors, for example, when the interferers are not transmitting signal or when the digital twin can provide accurate predictions, etc.
As a result, after the switching is finished, the reward signal required by the RL agent will be provided by the trained digital twin instead of the actual environment.
The agent keeps interacting with the digital twin until it does not improve, which marks the saturation of the agent learning and the end of the virtual interaction process.

\noindent\textbf{Demand based switching and active data acquisition:}
The system might require executing the above steps multiple times, based on the achieved performance.
The motivation of such repetition can be summarized as follows.
From the model training perspective, the quality of the collected datasets, i.e., $\mathcal{D}_{\mathrm{in}}$ and $\mathcal{D}_{\mathrm{s}}$, has significant influence on the prediction accuracy of the trained digital twin.
To be more specific, during the initial interaction process, most of the beams tried out by the agent are relatively random and hence have relatively poor quality in terms of SINR performance.
This means that the datasets are, intuitively speaking, biased towards the ``poor-quality'' beams.
As a result, the trained digital twin will have relatively inaccurate predictions on the beams that actually have better performance.
The incurred residual prediction error will in turn influence the learning of the agent, leading to unsatisfactory performance.

However, as the policy of the RL agent gets improved over time, the actions performed by the agent, i.e., the beams, are more likely to be in the beam space where the achieved SINR is high.
Therefore, it is advisable to switching back to the actual environment to re-collect data (through agent-environment interaction).
Such active data acquisition can enhance the training datasets with ``high-quality'' beams.
Using those better data samples to refine the parameters of the digital twin can help achieve higher prediction accuracy in the interested beam space, which further helps the learning of the agent.
By alternatingly performing these steps, the system has higher chance to collect data samples that are more useful for the agent learning, which has the potential of further enhancing both sample efficiency and learning convergence.
We show such interplay between the RL agent, actual environment and the digital twin in \fref{fig:surrogate-assisted}.

\section{Simulation Results} \label{sec:Simu}

\begin{figure*}[t]
	\centering
    \subfigure[Signal and interference prediction ($M=8$)]{ \includegraphics[width=0.305\textwidth]{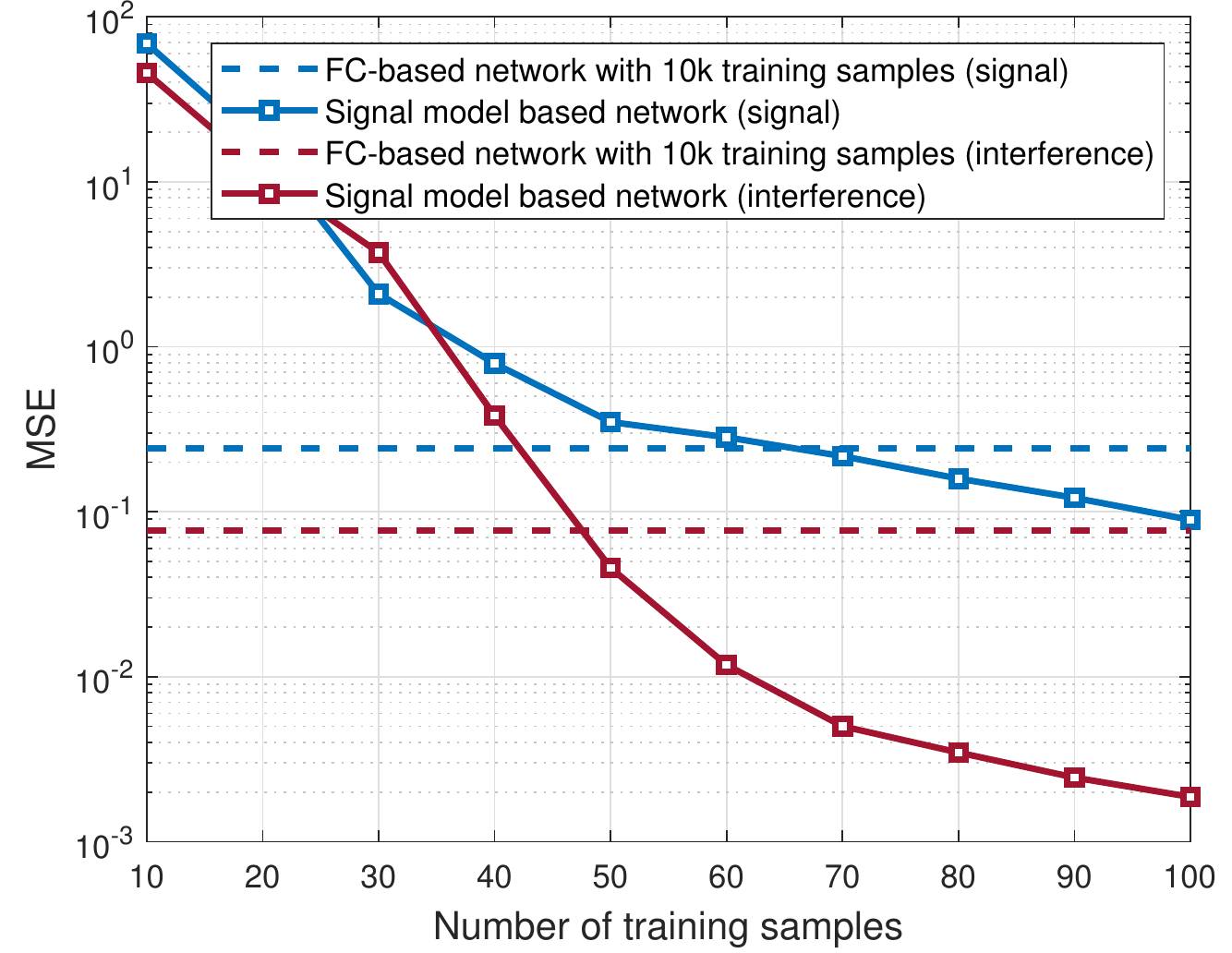} \label{acc-vs-sample-8} }
    \subfigure[Interaction with the actual environment]{ \includegraphics[width=0.32\textwidth]{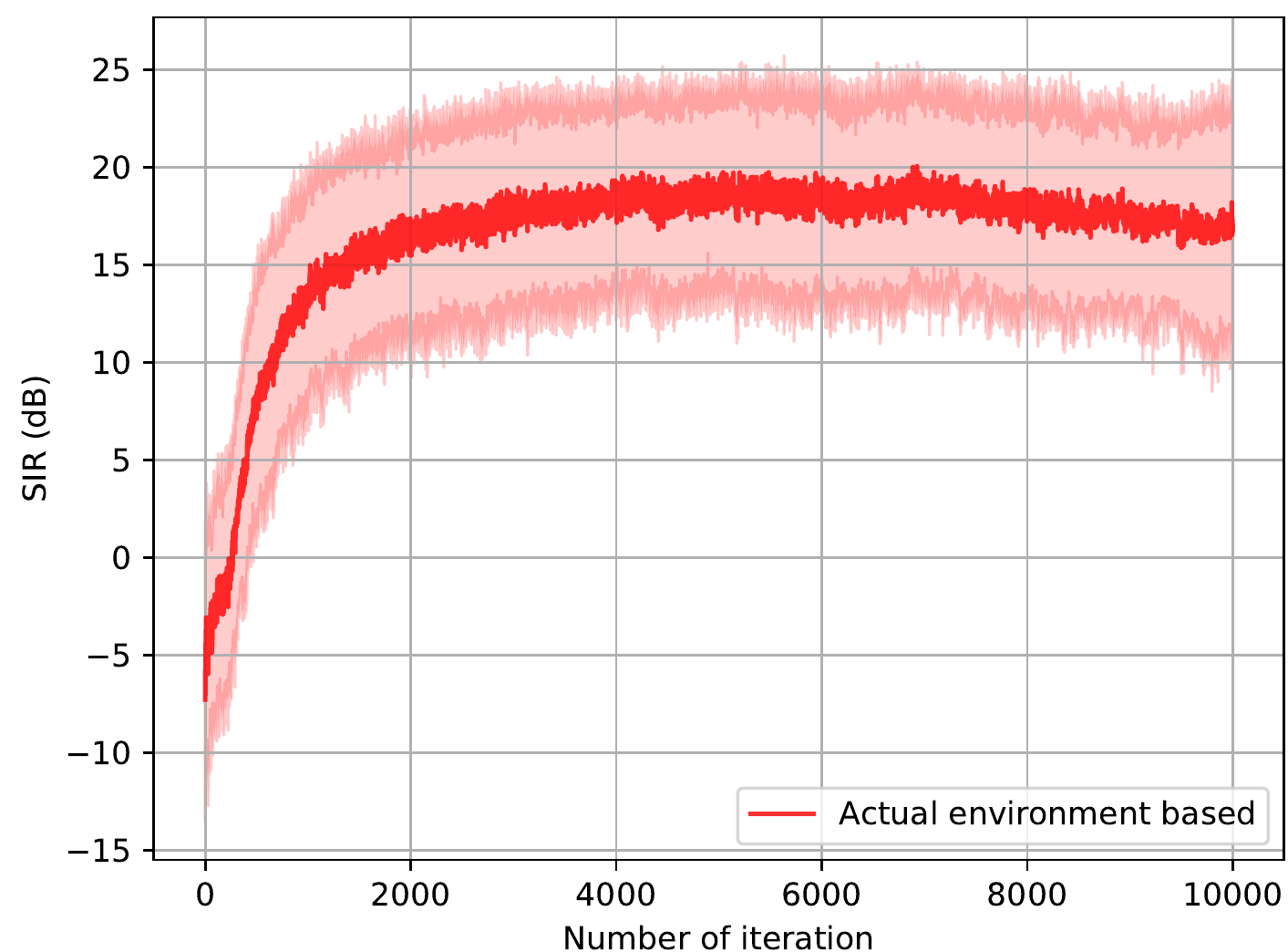} \label{actual-based} }
    \subfigure[Interaction with the digital twin]{ \includegraphics[width=0.32\textwidth]{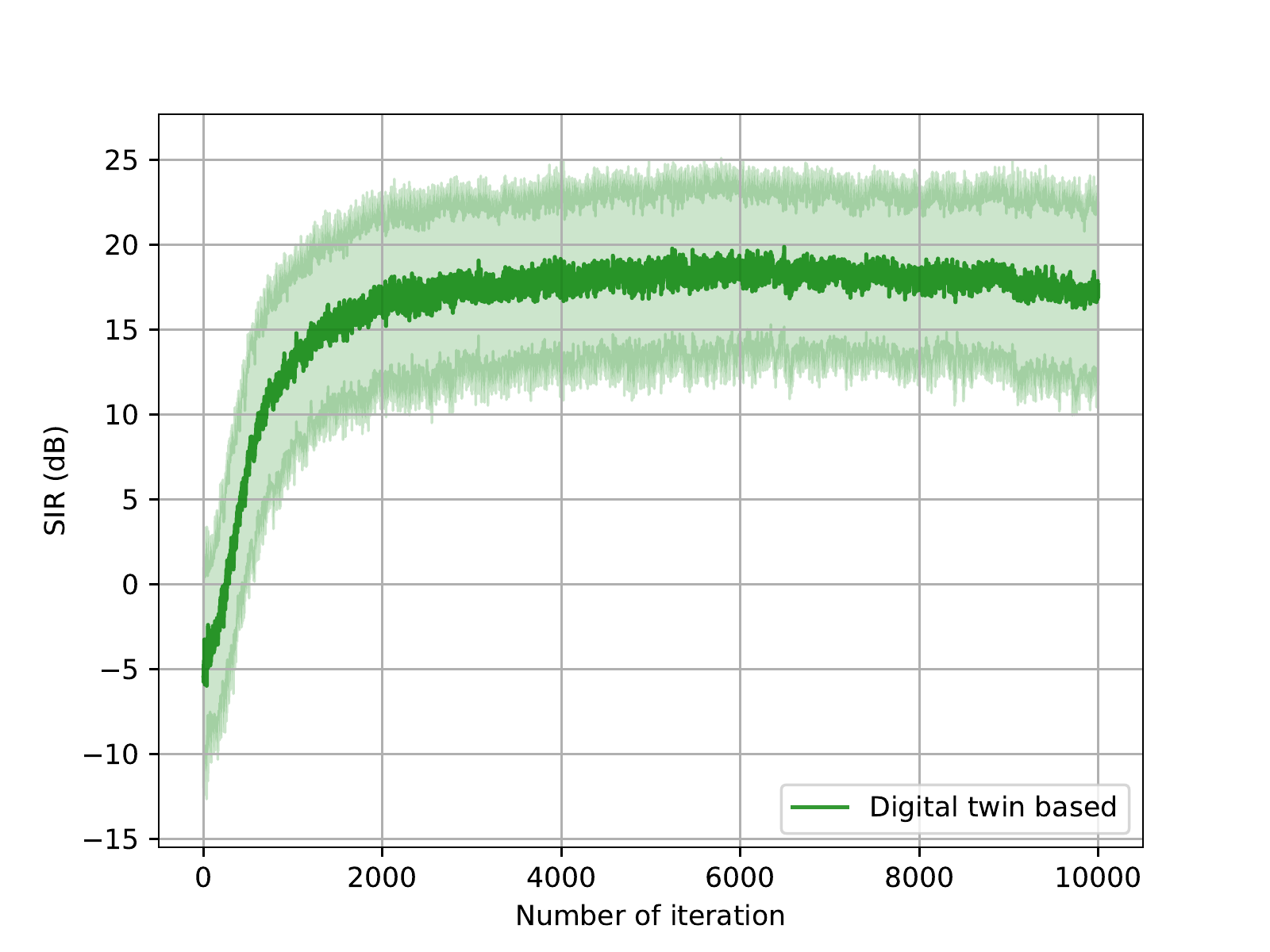} \label{surrogate-based} }
	\caption{The performance of the proposed digital twin-assisted beam learning framework. In (a), we compare the proposed signal model-based digital twin design with the FC-based design to show the significant reduction on the required real measurements. In (b) and (c), we show the learning experience of the DRL agent when interacting with the actual environment and the trained digital twin, to highlight the efficacy of such ``virtual'' environment.}
	\label{fig:iteract}
\end{figure*}

\subsection{Deep Learning Models and Training Procedures} \label{sub:DLA}

\subsubsection{DRL agent architecture}

Since the input of the actor network is the state and the output is the action, the size of both the input and output of the actor network is $M$, i.e., the number of antennas.
The critic network takes in the state-action pair and outputs the predicted Q value and hence it has an input size of $2M$ and an output size of $1$.
Both the actor and critic networks have two hidden layers in our proposed architecture, with the size of the first hidden layer being $16$ times of the input size and the size of the second hidden layer being $16$ times of the output size in both networks.
All the hidden layers are followed by the batch normalization layer for an efficient training experience and the Rectified Linear Unit (ReLU) activation layer.
The output layer of the actor network is followed by a Tanh activation layer scaled by $\pi$ to make sure that the predicted phases are within $(-\pi, \pi]$ interval.
The output layer of the critic network is a linear layer.
Moreover, we adopt the same DRL architecture for both solutions, regardless of having digital twin or not.

\subsubsection{Digital twin architecture}

We describe the two different architectures of the digital twin studied in this paper.
Also, as the signal prediction network and the interference prediction network have identical architecture in both solutions (i.e., model-based solution and fully-connected neural network based solution), for brevity, we only use the interference prediction network as an example.

\noindent\textbf{Signal model-based prediction network:}
As mentioned before in \eqref{in-net}, the interference prediction network is essentially devised to take a quadratic form of the combining vector determined by a positive semi-definite matrix $\mathbf{Q}_{\mathrm{in}}\mathbf{Q}_{\mathrm{in}}^H$, leaving the matrix $\mathbf{Q}_{\mathrm{in}}$ to be the model parameter.
Moreover, $\mathbf{Q}_{\mathrm{in}}$ has a shape of $M\times r_{\mathrm{in}}$ with $M$ being the number of antennas and $r_{\mathrm{in}}$ being a hyper-parameter.
The choice of $r_{\mathrm{in}}$ is empirically guided by the following rules: (i) $r_{\mathrm{in}}$ should not be too large as it will increase the model complexity and hence the required amount of training data; (ii) $r_{\mathrm{in}}$ should not be too small as it will limit the expressive capability of the model, leading to unsatisfactory prediction accuracy.

\noindent\textbf{Fully-connected neural network based prediction network:}
We adopt the fully-connected neural network with two hidden layers to be the interference prediction network.
The input layer of the network has $M$ neurons, which is equal to the number of antennas.
The output layer of the network has only one neuron with linear activation.
Both hidden layers have $M^\prime$ neurons.
Similar to $r_{\mathrm{in}}$ in the model-based architecture, the selection of $M^\prime$ needs to strike a balance between model complexity and model expressive capability.
Moreover, all the hidden layers are followed by the batch normalization layer and ReLU activation layer.

\begin{table}[t]
\caption{Hyper-parameters for digital twin training}
\centering
\begin{tabular}{c|c|c}
  \hline
  \hline
  \textbf{Parameter} & \textbf{Model-based} & \textbf{FC-based} \\
  \hline
  Batch size & 512 & 512 \\
  Number of epochs & 500 & 500 \\
  Optimizer & Adam & Adam \\
  Initial learning rate & $1\times10^{-1}$ & $1\times10^{-2}$ \\
  Learning rate schedule & $0.1$@$\left[50, 300, 400\right]$ & $0.1$@$\left[100, 300, 400\right]$ \\
  \hline
  \hline
\end{tabular}
\label{TrParam}
\end{table}

\subsubsection{Training parameters}

As mentioned before, the digital twin is trained in a supervised fashion, based on the collected power datasets, i.e., $\mathcal{D}_{\mathrm{in}}$ and $\mathcal{D}_{\mathrm{s}}$.
Moreover, the interference prediction network and the signal prediction network are independently trained.
However, for the same type of digital twin, i.e., either model-based or fully-connected neural network based, we adopt the same training parameters for interference and signal prediction networks.
We summarize the detailed hyper-parameters used for training the digital twins in \tref{TrParam}.

\subsection{Numerical Results} \label{subsub:NResult}

In this subsection, we provide the simulation results of the proposed digital twin-assisted interference-aware beam learning solutions.
We first evaluate the prediction accuracy of the two proposed prediction network architectures, which provides insight on how much data samples are required in order to have a reasonable performance as well as the practicality of the solutions.
We show the prediction accuracy of both the signal power and the interference power.
As can be seen, the signal model-based architecture requires much less data samples to achieve higher prediction accuracy than the FC-based architecture trained with much more data samples.
For instance, as indicated in \fref{acc-vs-sample-8}, with only $50$ samples, the signal model-based prediction architecture can achieve even more accurate interference prediction than the FC-based architecture trained with $10,000$ samples.
\textbf{This saves almost $99.5\%$ of the measurements, yielding a more sample-efficient solution for the practical system deployment.}
Moreover, as there are more data samples, the prediction accuracy of the signal model-based architecture also gets improved quite significantly.
Such performance is achieved by better leveraging the underlying signal relationships and hence the model parameters are essentially searched over a much smaller space.
The trained digital twin is utilized to interact with the DRL agent, aiming to reduce the expensive actual measurements conducted by the hardware.
In \fref{fig:iteract}, we show the performance of the DRL agent when interacting with the actual environment as well as the digital twin.
The training of the DRL agent is repeated for $100$ times and the average performance as well as the standard deviation are reported in \fref{fig:iteract}.
We test the performance of a system with $8$ antennas, and the digital twin is trained using $1,000$ data samples, i.e., $|\mathcal{D}_{\mathrm{in}}|=|\mathcal{D}_{\mathrm{s}}|=1000$.
As can be seen, the learning experience based on the digital twin is quite similar to that of the one based on the actual environment.
This empirically shows the effectiveness of using the digital twin in training the DRL agent.
As a result, although the DRL agent requires almost a total number of $5,000$ interactions with the environment to converge, \textbf{in the digital twin assisted learning framework, all these interactions are with the digital twin and hence the expensive evaluations on the real hardware are avoided.}

\section{Conclusion} \label{sec:Con}

In this paper, we developed a sample-efficient digital twin-assisted online interference-aware beam design framework.
The proposed solution learns how to design beam patterns that can effectively manage interference, relying only on the power measurements and without any channel knowledge.
The design of the digital twin leverages the underlying signal relationship, leading to a significant reduction on the required interactions with the actual environment.
Moreover, it also facilitates other tasks such as interference-aware codebook learning, where the data sharing among different beam learning agents/engines becomes possible.
The results highlight the efficacy of the trained digital twin in guiding the beam learning process.
%


\end{document}